\documentclass[aps,twocolumn,pre]{revtex4-1}
\usepackage{graphicx}
\usepackage{amsmath,mathtools}
\usepackage{epstopdf}
\usepackage{color}

\begin{document}

\title{Nonequilibrium fluctuations of a driven quantum heat engine via machine learning}
\author{Sajal Kumar Giri$^a$ and Himangshu Prabal Goswami$^{a,b}$}
\email{hpgoswami@pks.mpg.de}
\affiliation{$^a$Finite Systems Division, Max-Planck-Institute for the Physics of Complex Systems, Dresden-01187, Germany}
\affiliation{$^b$Department of Chemical Sciences, Tezpur University, Napaam, Tezpur-784028, Assam, India}

\date{\today}

\begin{abstract} 
We propose a machine learning approach based on artificial neural network to gain faster insights on the role of geometric contributions to the nonequilibrium fluctuations of an adiabatically temperature-driven quantum heat engine   coupled to a cavity. Using the artificial neural network we have explored the interplay between bunched and antibunched photon exchange statistics  for different engine parameters. We report that beyond a pivotal cavity temperature, the Fano factor oscillates between giant  and low values as a function of phase difference between the driving protocols. We further observe that the standard thermodynamic uncertainty relation is not valid when there are finite geometric contributions to the fluctuations, but holds true for zero phase difference even in presence of coherences.
\end{abstract}

\maketitle

\section{ Introduction}
Random fluctuations of an observable are ubiquitous in the statistical analysis of nonequilibrium and open quantum systems.  A standard tool to quantify fluctuations associated with heat or particle  transport in small systems is the Full Counting Statistics (FCS) approach \cite{belzig2005full, nazarov, Levitov, uhrmp}. It allows understanding of the underlying distribution functions by evaluating the moments and cumulants to all orders \cite{ uhrmp,campisi2011colloquium, dharheat}.  Statistical studies based on FCS led to the development of universal nonequilibrium fluctuations theorems and also have recently led to the development of thermodynamic uncertainty relationship \cite{barato2015thermodynamic, pietzonka2017finite, proesmans2017discrete,horowitz2017proof}, strengthening the principles of quantum thermodynamics\cite{dubi2011colloquium,gustavsson2006counting,SaarQHEPhysRevA.86.043843}. The principles of FCS has also been recently used to understand universal oscillations in higher order time-dependent cumulants \cite{flindt2009universal} as well as experimentally verify fluctuation theorems in a bidirectional electron counting device \cite{utsumi2010bidirectional}. On the theoretical front, FCS is particularly useful because it provides an analytical method to study the statistics when combined with a generating function technique based on either master equations or nonequilibrium Green's functions\cite{max-uh,wang2014nonequilibrium,uhrmp}. 
 
 FCS of open quantum systems with several manybody states
 is a challenging problem when analyzed via a master equation technique.  Under this scenario, often analytical expressions for the generating function cannot be derived and one has to resort to numerics to evaluate the moments and cumulants. Understanding parameter dependences on the moments and cumulants through numerical analysis  becomes extremely time consuming when one has several parameters to scan for, eg. in quantum heat engines\cite{kosloff2013quantum,xu2016polaron,SaarQHEPhysRevA.86.043843} 
 and multilevel quantum dot or single molecule junctions\cite{uhFTfcs}. This problem is further aggravated when a few parameters of the system are modulated in time. It has been recently shown in both electron and heat transport that externally driving the temperature of reservoirs results in  geometric augmentations (Pancharatnam-Berry phaselike (PBp) effects) to the cumulants  when evaluated via the master equation method \cite{ren2010berry, PhysRevA.95.023610,hayakawa,PhysRevE.96.022118,PhysRevA.95.023610}. Such geometric contributions or PBp effects have  strange ramifications on the overall statistics, for example,  violation of the standard mathematical nonequilibrium fluctuation relations \cite{ren2010berry,hpg4,PhysRevE.96.022118,PhysRevE.96.052129}. 

 
 In this work, we propose a machine learning technique
 based on artificial neural network (ANN)\cite{dalton1991artificial,carleo2017solving,PhysRevLett.98.146401,PhysRevLett.119.150601,Hush580,PhysRevB.97.035116} to bypass the effort associated with such time-consuming numerics. Study of open quantum system using machine learning is a growing field of research \cite{schuld2015introduction,biamonte2017quantum}. Recently  the exciton dynamics of photosynthetic complexes has been studied using multi-layer perceptrons \cite{C5SC04786B} and deep learning \cite{C7SC03542J}. Machine learning algorithms have also been used to model electronic transmission coefficients (Greens functions) in molecular junctions\cite{PhysRevB.89.235411}. A type of ANN is the  feedforward protocol which  works on the principle of
  backpropagation of error and the generation of a  linear or nonlinear mapping between input and output data.  Here, we report that ANN can be effectively used to regressively evaluate the moments and cumulants to a very good precision under supervised learning.  To the best of our  knowledge this is the first application of machine learning tools in understanding parameter dependence on the FCS of open quantum systems.

Recently, we showed that in a $4$ level driven quantum heat engine (dQHE) coupled to a unimodal cavity, there exists a competition between thermally induced quantum coherences and the geometric contributions such that the latter stops  the former to optimize the flux into the cavity mode when the temperature of the two thermal reservoirs are periodically modulated in time in an adiabatic fashion\cite{PhysRevE.96.052129}. In this work, we focus on the geometric contributions to the higher order nonequilibrium fluctuations (noise) with a view to further understand the role of the geometric effects on the statistics. We analyze the geometric contributions to the statistics in terms of the Fano factor ($F$) by calculating the ratio of the first and second order cumulants (variance to mean ratio). $F$ is the measure of bunched ($F>1$) and antibunched ($F < 1$) photon exchange statistics (PES) \cite{PhotonAntibunchingRevModPhys.54.1061}. $F$ is also one of the the central quantities in the  universal thermodynamic uncertainty relationship along with the thermodynamic affinity which quantify a trade-off between  the rate of entropy production and noise \cite{koyuk2018generalization}. Recently, there have been claims that quantum coherences can break this relationship during heat and electron transport\cite{PhysRevB.98.085425, agarwalla2018}. In this work, $F$ is the quantity under investigation using supervised learning based  ANN with a motive to understand parameter dependences such as coherences and PBp effects on the photon exchange statistics and the thermodynamic uncertainty relationship.

The paper is organized as follows. In Sec.(\ref{qhe}), we briefly review the essentials of the dQHE, FCS and PBp contributions. In Sec.(\ref{ann}),  we discuss the ANN modelling of the dQHE's output. We use the ANN to study the photon exchange statistics (PES) and the thermodynamic uncertainty relationship in Sec.(\ref{stats}) and (\ref{tur}) respectively.  We finally conclude in Sec.(\ref{conc}).

\begin{figure}
\centering
\includegraphics[width= 0.5 \textwidth]{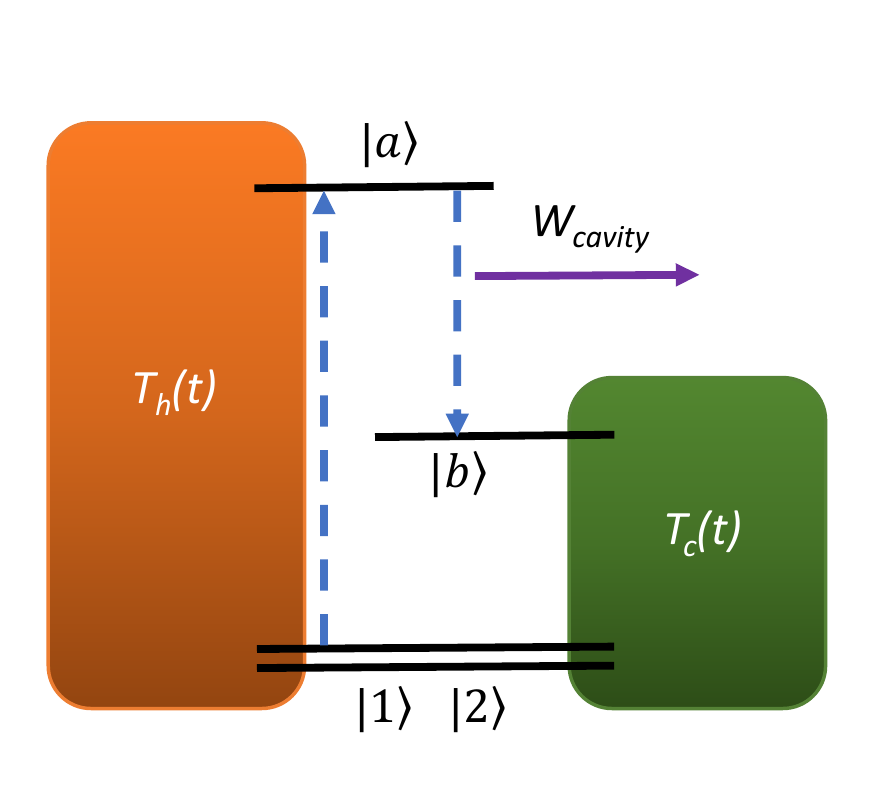}
\caption{Schematic plot of a dQHE with $4$ energy levels. Two degenerate states $| 1 \rangle$ and $| 2 \rangle$  are coupled with higher energy states $| a \rangle$ and $| b \rangle$ through thermal baths. Hot and cold bath temperatures are labeled as $T_h(t)$ and $T_c(t)$ respectively. States $| a \rangle$ and $| b \rangle$ are also coupled to a unimodal cavity when emission of photons is the work done.}
   \label{qhe_schematic}
\end{figure}

\section{Driven quantum heat engine}
\label{qhe}

We consider a $4$ level QHE coupled asymmetrically to two thermal baths and an unimodal cavity, Fig.(\ref{qhe_schematic}). This model has been studied in several works \citep{PhysRevE.96.052129,SaarQHEPhysRevA.86.043843,UHeplQHE,goswami2013thermodynamics,ScullyPNAS13092011}. The model consists of two thermal baths at temperatures $T_h(t)$ and $T_c(t)$. We express thermal bath temperatures as $T_c(t)=(T_{c0}-A_0\sin(\omega t))$ and $T_h(t)=(T_{h0}-A_0\sin(\omega t +\phi))$, where 
$A_o, \omega$ and $ \phi$ are the amplitude, frequency and phase difference between the driving protocols respectively. The cold (hot) bath temperature oscillates with frequency $\omega$ around the initial value $T_{c0}$($T_{h0}$). Bath temperatures are periodically driven in time such that, $T_h(t)>T_c(t)$ condition is maintained throughout. Two degenerate quantum states $| 1 \rangle$ and $| 2 \rangle$, with same symmetry (therefore with a forbidden transition between them) are coupled to two thermal baths. The higher energy states $| a \rangle$ and $| b \rangle$ with different symmetry and allowed transition between them are coupled to the hot and cold bath respectively. The state $| a \rangle$ is higher in energy than the state $| b \rangle$. $| 1 \rangle$, $| 2 \rangle$, $| b \rangle$ and $| a \rangle$ states correspond to the energies of $E_1$, $E_2$, $E_b$ and $E_a$ respectively. We assume that all the couplings between system and thermal bath are equal and denoted by $r$. States $| a \rangle$ and $| b \rangle$ are also coupled to a unimodal cavity and the strength of the coupling is denoted by $g$. With above assumptions the total Hamiltonian can be written as $\hat{H}_T= \hat{H}_o+\hat{V}+\hat V^\dag$, where
\begin{eqnarray}
\label{Hamiltonian}
\hat{H}_o&=&\displaystyle\sum_{\nu=1,2,a,b}
E_\nu|\nu \rangle\langle \nu^\prime |+\displaystyle\sum_{k\in h,c}\epsilon_k^{}\hat{a}_k^\dag\hat{a}
_k+\epsilon_l\hat{a}_l^\dag\hat{a}_l,\\
\hat V&=&\displaystyle\sum_{k\in h,c}\sum_{i=1,2}\sum_{x=a,b}r\hat{a}_k|x\rangle\langle i|+g(\hat{a}_l^\dag|b\rangle\langle a|+|a\rangle\langle b|\hat{a}_l)\nonumber .\\
\end{eqnarray}
In the above equation, $E_\nu$, $\epsilon_k$ and $\epsilon_l$ are the energy of the system's $\nu$th level, $k$th mode of the thermal reservoirs and unimodal cavity respectively. Thermal baths are modeled as harmonic modes with $\hat{a}^\dag (\hat{a})$ being the bosonic creation (annihilation) operators. There is a heat flow from the hot bath to the cold bath in a nonlinear fashion. A  radiative decay channel originates from the transition $|a\rangle\rightarrow|b\rangle$ and is coupled to a unimodal cavity. 

We will now proceed to quantify the PES between system and cavity.
We employ an adiabatic Markovian quantum master equation approach combined with a standard generating function technique to evaluate the statistics such that $|\dot\rho(\lambda,t)\rangle = \breve{{\cal L}}(\lambda,t)|\rho(\lambda,t)\rangle$, where  $\lambda$ is a field that counts the number of photon exchanged between system and cavity \cite{PhysRevE.96.052129}. $|\rho(\lambda,t)\rangle = \{\rho_{11},\rho_{22},\rho_{aa},\rho_{bb},\Re(\rho_{12})\}$ is the reduced system density vector with $\rho_{ii}, i=1,2,a,b$ being system's many body states and $\Re(\rho_{12})$ is the thermal induced coherences between states $| 1 \rangle$ and $| 2 \rangle$. $\breve{{\cal L}}(\lambda,t)$ is the adiabatic effective evolution Liouvillian superoperator within the Markov approximation. The statistics of $q$ (number of photons exchanged between the system and the cavity) is obtained from  moment generating function, which is expressed as $G(\lambda,t)=\sum_qe^{\lambda q}P(q,t)$ where $P(q,t)$ is  the probability distribution function corresponding to $q$ net photons in the cavity  within a measurement window, $t$. Within  the FCS formalism, it can be shown that $\dot G(\lambda,t)=\langle \breve{{\boldsymbol 1}}| \breve{\mathcal L}(\lambda,t)|\rho(\lambda,t)\rangle$ with $\langle\breve{\bf 1}|=\{1,1,1,1,0\}$ \cite{Levitov,uhrmp}. The full form of the characteristic Liouvillian is given by \cite{PhysRevE.96.052129}, $\breve{\mathcal L}(\lambda,t)=$ \begin{equation}
 \label{Liouvillian_lambda}
r\begin{pmatrix}
n(t)&0&\tilde{n}_h(t)&\tilde{n}_c(t)&y(t)\\
0&n(t)&\tilde{n}_h(t)&\tilde{n}_c(t)&y(t)\\
n_h(t)&n_h(t)&\frac{-g^2\tilde{n_l}-2r\tilde{n}_h^{}(t)}{r}&\frac{g^2n_l e^{-\lambda}}{r}&2p_hn_h(t)\\
n_c(t)&n_c(t)&\frac{g^2\tilde{n_l}e^{\lambda}}{r}&\frac{-g^2n_l-2r\tilde{n}_c(t)}{r}&2p_cn_c(t)\\
\frac{y(t)}{2}&\frac{y(t)}{2}&p_h\tilde{n}_h(t)&p_c\tilde{n}_c(t)&n(t)
\end{pmatrix}.\\[2mm]
\end{equation}
In the above equation  $n(t)=-(n_c(t)+n_h(t))$, $y(t)=-(n_c(t)p_c+n_h(t)p_h)$, $\tilde{n}_c(t)=n_c(t)+1$,  $\tilde{n}_h(t)=n_h(t)+1$ and $\tilde n_l=1+n_l$. The explicit form of $n_c(t)$, $n_h(t)$ and $n_l$ can be expressed as $ n_c(t)=1/(\exp\{(E_b-E_1)/k_BT_c(t)\}-1), n_h(t)=1/(\exp\{(E_a-E_1)/k_BT_h(t)\}-1)$ and $n_l=1/(\exp[(E_a-E_b)/k_BT_l]-1)$. Here, $T_l$ is the fictitious temperature of the cavity,  $p_h$ and $p_c$ are quantum coherence control parameters associated with the hot and cold  baths respectively \cite{ScullyPNAS13092011}.

Within the FCS formalism, in the long time limit, one can obtain the PBp contributions from scaled cumulant generating function given by $S(\lambda)=\lim_{t\rightarrow\infty}(1/t)\ln[\langle\breve{\bf 1}|\exp(\breve{\cal L}(\lambda,t)t)|\rho(\lambda,t)\rangle]$. $S(\lambda)$ is additively separable into a dynamic and a geometric part, $S(\lambda)=S_d(\lambda)+S_g(\lambda)$ \cite{sin} ,
 \begin{eqnarray}
 \label{s-dyn}
S_d(\lambda)&=&\frac{1}{t_p}\displaystyle\int_0^{t_p}\zeta_o(\lambda,t')dt',\\
S_g(\lambda)&=&-\frac{1}{t_p}\int_0^{t_p}\langle L_o({\lambda,t'})
 |\dot R_o({\lambda,t'})\rangle dt'.
 \label{s-geo1}
 \end{eqnarray}
In the above equation, $S_{d}(\lambda)$ and $S_{g}(\lambda)$ represent the dynamic and geometric cumulant generating function respectively. $|R_o(\lambda,t')\rangle $ and $\langle L_o(\lambda,t')|$ are the instantaneous right and left eigenvectors of $\breve{\cal L}(\lambda,t')$ with instantaneous long-time dominating eigenvalue, $\zeta_o(\lambda,t')$. Here $t_p$ ($2\pi/\omega$) is the driving period. Note that, analytical expressions for both $S_d(\lambda)$ and $S_g(\lambda)$ cannot be derived for this $4$ level dQHE. Cumulant generating functions within the adiabatic master equation are analytically known only for two level systems\cite{ren2010berry,hpg4} within the Markov limits. For systems with large number of  states, analytical expressions have not been reported since the geometric contributions involve calculation of both the left and right eigenvectors of the Liouvillian. The $n$th order fluctuations (cumulants of $S(\lambda)$) can be calculated as,
\begin{align}
\label{cd}
C_d^{(n)}&=\partial_\lambda^n S_d(\lambda)|_{\lambda=0},\\
 \label{cg}
C_g^{(n)}&=\partial_\lambda^n S_g(\lambda)|_{\lambda=0}.
\end{align}
We will focus only on $n=1,2$ to get,
\begin{align}
\label{Fano}
F=\frac{c_d^{(2)} + c_g^{(2)}}{c_d^{(1)} + c_g^{(1)}},
\end{align}
where $c_d^{(1)} (c_g^{(1)})$ and $c_d^{(2)} (c_g^{(2)} )$ are first and second order dynamic (geometric) cumulants respectively.

\begin{figure}
\centering
\includegraphics[width= 0.5 \textwidth]{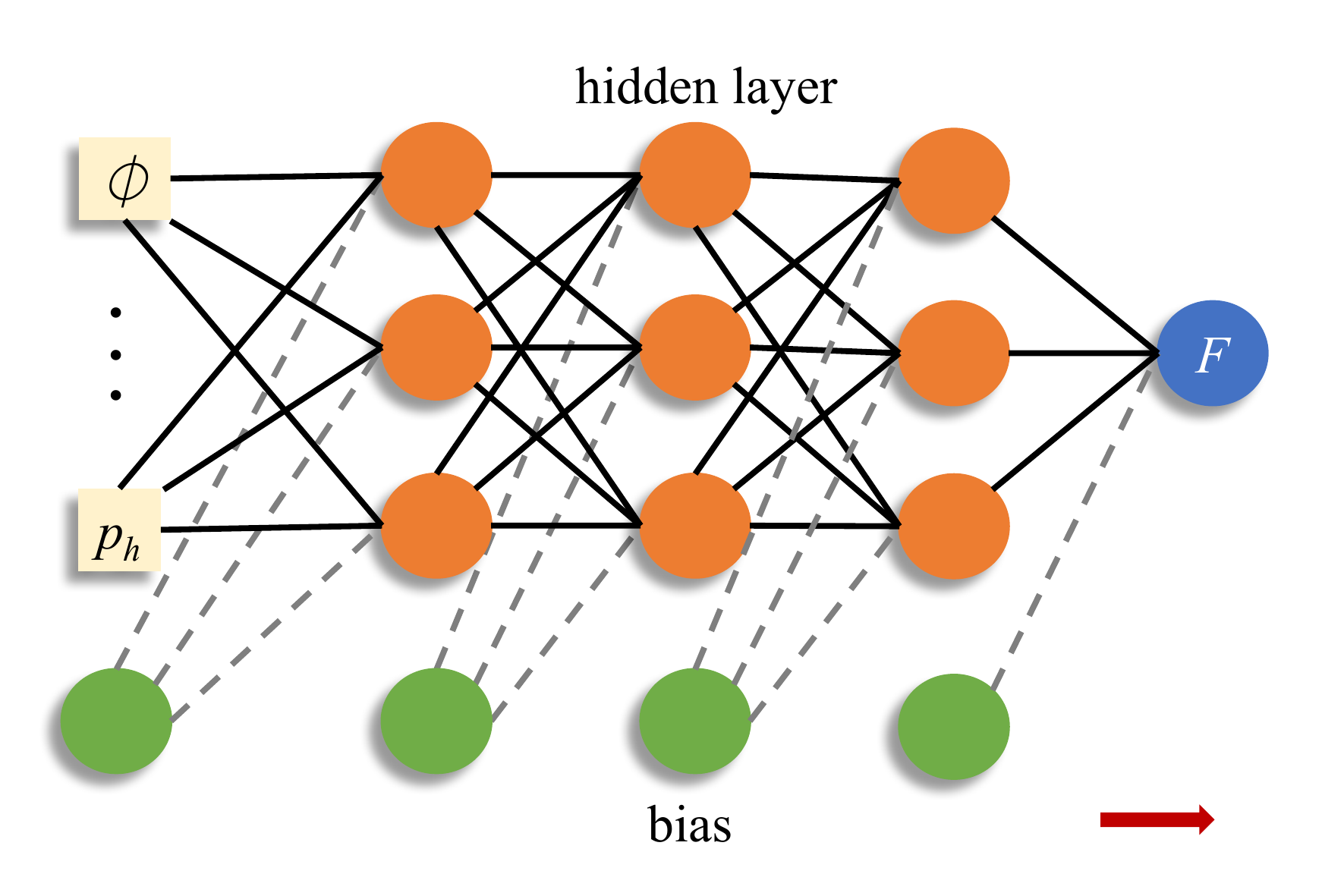}
\caption{Schematic plot of a fully connected feedforward artificial neural network (with $3$ hidden layers each containing $3$ neurons) showing a mapping from the Liouvillian parameters (input of the network) to $F$ (output of the network).  Orange circles represent hidden layer neurons and green circles represent biases linked to each hidden layer neurons and output layer.}
 \label{network}
\end{figure}

\section{FCS via machine learning }
\label{ann}
We will analyze the role of system parameters on the Fano factor, $F$, with the help on an ANN. In our formalism, the input of the network are parameters from the effective superoperator $\breve{\mathcal L}(\lambda,t)$ and the output is $F$. The mapping is continuous and we are interested in a regressive analysis. For regression,  ANN approximately develops a linear or non-linear functional mapping from the  input space to the output space.  In this work we have considered a supervised learning based ANN i.e. learning form labelled data to explore the photon fluctuation statistics of the dQHE. To establish and generalize the mapping, we split the full data in three parts, training (to establish the mapping), validation (to generalize the mapping) and test (to predict for test data). Within the general framework of ANN, data pairs (pairs of input and output data) are either obtained from  experiments or numerical simulations. Here, we generated data numerically by solving the Markovian quantum master equation with the analytically derived effective evolution operator given in Eq.(\ref{Liouvillian_lambda}). 

\begin{figure}
\centering
\includegraphics[width= 0.4 \textwidth]{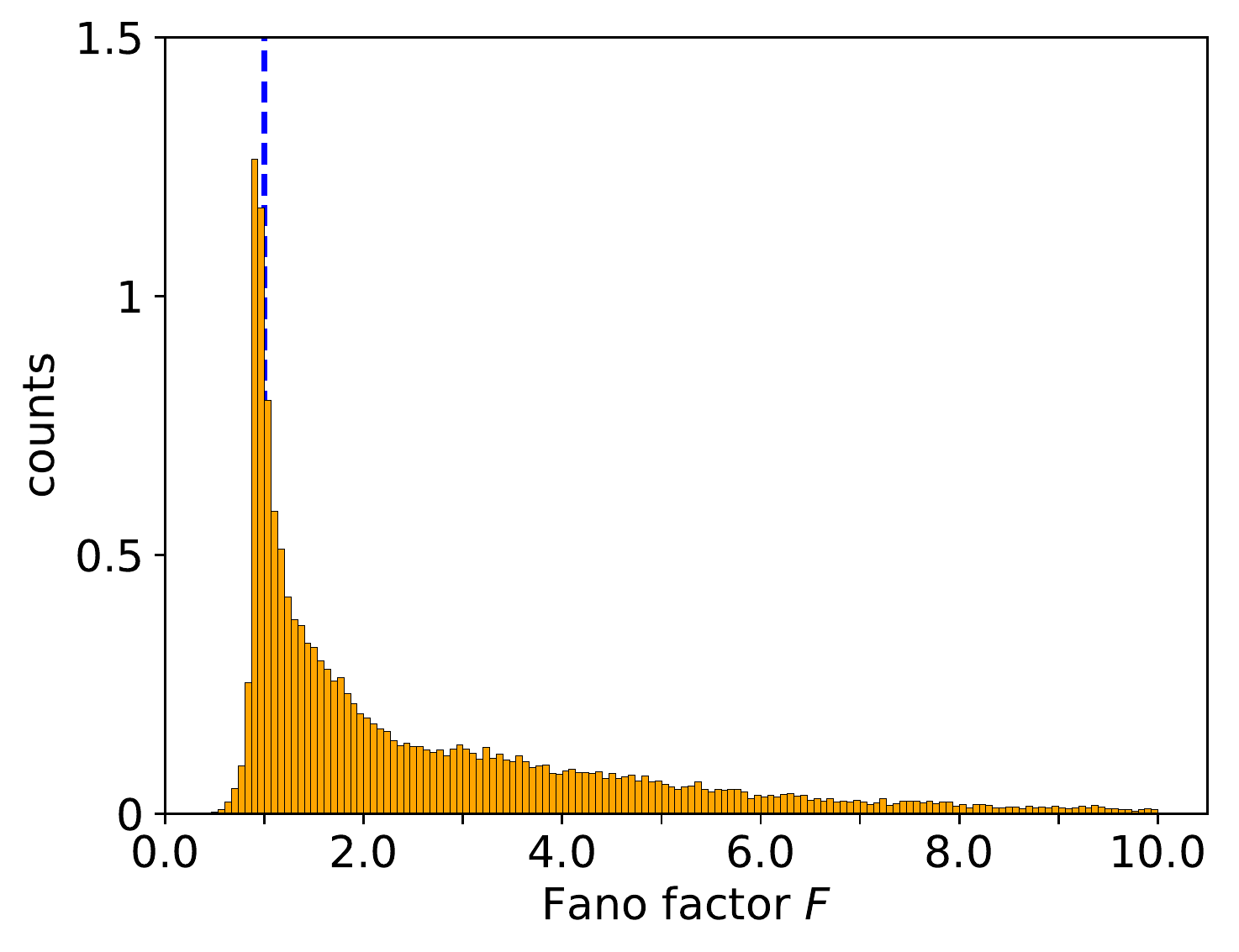}
\caption{Normalized histogram of the $F$ for considered parameter space of the Liouvillian. Blue dotted line represents $F=1$.}
   \label{fano_dist}
\end{figure}

To  explore the dependence of Liouvillian parameters on the Fano factor, we treat the former  as input and the latter as output of the network as shown schematically in Fig.(\ref{network}). For simplicity,  we consider only $6$ parameters from the Liouvillian as input of the network and these are $T_{c0}$, $T_{h0}$, $T_l$, $\phi$, $p_h$ and $p_c$. The range of Liouvillian parameters that have been used to generate uniformly distributed $30000$ random combinations are listed in Table.(\ref{param_table}). We have used the atomic unit unless specified otherwise. Note that the dynamic cumulants do not depend on $A_0$ and $\omega$, whereas the geometric cumulants change linearly with these two parameters and are hence kept fixed at $A_0=0.007$ and $\omega=0.7$ MHz. Other parameters are fixed at $E_1=E_2=0.1$, $E_b=0.3$, $E_a=1.5$, $r=5$ and $g=10$. The normalized histogram of the $F$ obtained from the numerical simulation (for $30000$ uniformly distributed random selection of $6$ Liouvillian parameters), is shown in Fig.(\ref{fano_dist}). The histogram peaks at around $F=0.93$ and the average value of $F$  is $2.58$. There are around $80\%$ data for $F>1$, around $18\%$ data for $F<1$ and very small number of data (around $2\%$) for $F=1$.  

\begin{table}
\centering
\caption{Range of Liouvillian parameters.}
\label{param_table}
\begin{tabular}{ c c} 
 \hline
 \hline
parameters & value \\
\hline
$T_{c0}$ & $0.2-0.7$\\ 
$T_{h0}$ & [$T_{c0}+0.5] -[T_{c0}+1.2$] \\ 
$T_l$ & $0.1-1$  \\
$\phi$ &  $0-2\pi$ \\
$p_h$ & $0-1$ \\
$p_c$ & $0-1$ \\

 \hline
 \hline
\end{tabular}
\end{table}

In the ANN, the input layer is connected with the output layer through some hidden layers as shown in Fig.(\ref{network}). The hidden layer contains arbitrary number of neurons with activation functions and performs a transition from a previous  layer to the next. The function of the output layer neuron is known as transfer function. The activation and transfer functions can be linear or nonlinear. In our case, we have used the $tanh$ function as the hidden layer activation function and the $pure \ linear$ function ($y=x$) as the transfer function. The output from the individual neurons and the bias elements in $l$th layer is represented by the vectors $\vec X_l$ and $\vec B_l$ respectively. Any $l$th layer is connected to the next $(l+1)$th layer via a weight matrix ${\bf W}_l$ in following way,
\begin{equation}
 \vec X_{l+1}=\tanh[{\bf W}_l^{T}\vec X_l + \vec {B_l}].
\end{equation}
In the above equation ${\bf W}_l^{T}$ is the transpose of ${\bf W}_l$. The dimension of $\vec X_l$, ${\bf W}_l$ AND $\vec B_l$ are $v_l \times 1$, $v_l \times v_{l+1}$ and $v_{l+1} \times 1$ respectively, where $v_{l+1}$ ($v_{l}$) is the number of neurons in $(l+1)$th ($l$th) layer. Mean squared error ($MSE$) is used as a cost function for the model evaluation. The coefficient of determination ($R^2$) is also used as another indicator to determine the network performance. $MSE$ and $R^2$ can be expressed as 
\begin{eqnarray}
\label{mse_r2}
MSE&=&\frac{1}{N}\sum_i (F_i^{true}-F_i^{pred})^2, \\
R^2&=&1-\frac{\sum_i (F_i^{true}-F_i^{pred})^2}{\sum_i (F_i^{true}-\bar{F}^{true})^2},
\end{eqnarray}
where $N$ is the number of examples or data pairs, $F_i^{true(pred)}$ is the actual (predicted) value of  $F$ of $i$th example and $\bar{F}^{true}$ is the averaged actual value of $F$. Note that $R^2$ is a unitless quantity and bounded between $-\infty$ and $1$. Root mean squared error ($RMSE$) is the squared root of  $MSE$ and has the same unit as the output (in present work it has the same unit as $F$ i.e dimensionless). For the best prediction, value of $RMSE$ ($R^2$) is zero (one). We used the Levenberg-Marquardt (LM) algorithm\cite{haykin2009neural} for the backpropagation of errors. The LM algorithm  updates weights and biases in following fashion,
\begin{equation}
\label{coef_update}
\vec z_{i+1}=\vec z_i-[{\bf J}^T{\bf J} + \sigma{\bf I}]^{-1}{\bf J}^T \vec{e},
\end{equation}  
where ${\bf J}$ is the Jacobian matrix and it's elements are first derivatives of the ANN errors with respect to the weights and biases with ${\bf J}^T$ being the transpose. $\vec z_i-$s  represent the weights and biases after $i$th iteration and $\vec{e}$ is a vector  containing the ANN errors. Here ${\bf I}$ represents identity matrix. $\sigma$ is a control parameter that mediates the interplay between the Newton and the gradient descent method.  For $\sigma = 0$, Eq.(\ref{coef_update}) reduces to the Newton's method and for large $\sigma$, Eq.(\ref{coef_update}) is the gradient descent method. The value of $\sigma$ during the optimization can be adapted. We reduce the value of $\sigma$ by a factor $\sigma_d$ after each successful iteration since the Newton's method is faster and very effective near the minimum of error. $\sigma$ is increased by a factor $\sigma_i$ only when a step increases the performance error. Therefore the performance error always reduces at every iteration. If the iteration needs to stop early, we check the validation during the training with a patience of $6$ i.e. the training stops if the validation performance increases for $6$ epochs in a row. 

\begin{figure}
\centering
\includegraphics[width= 0.5 \textwidth]{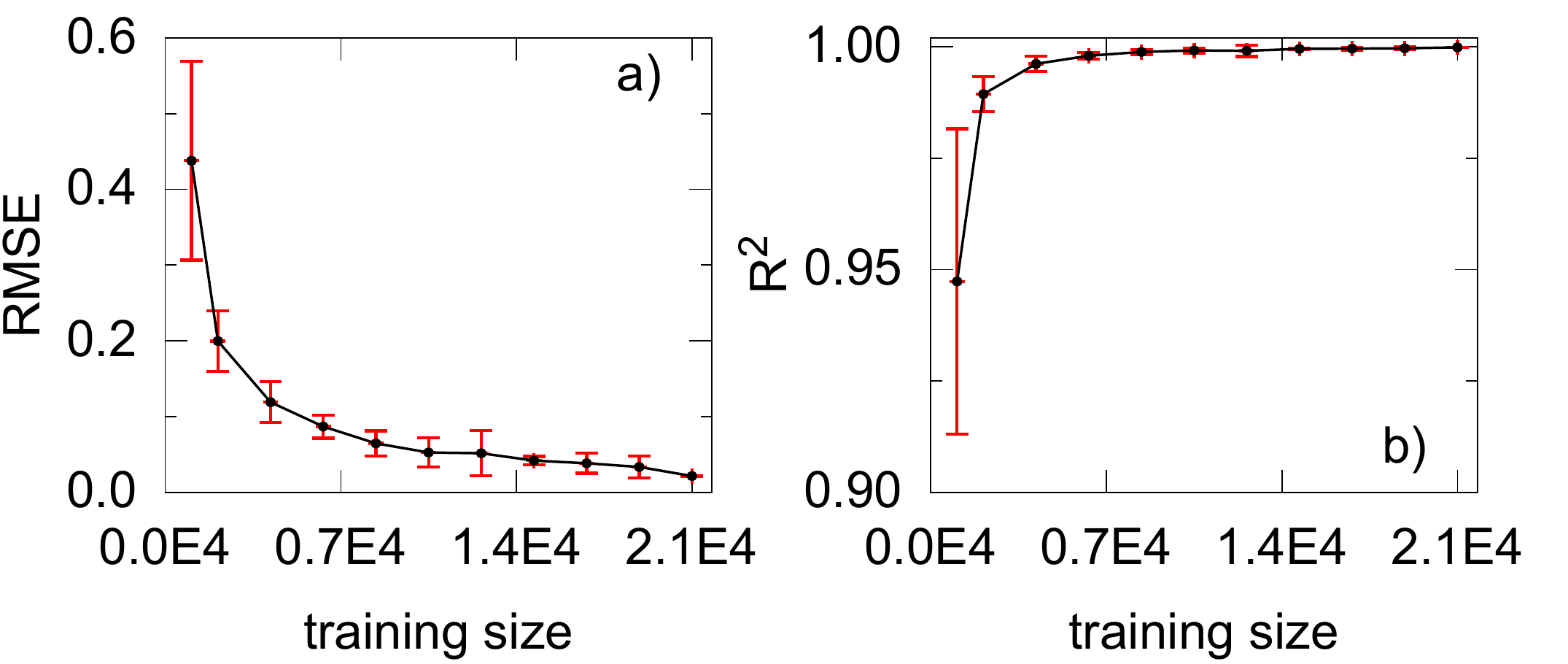}
\caption{$RMSE$ (a) and $R^2$ (b) against training size for test data only. Error bars are obtained from $10$ trials. }
   \label{error}
\end{figure}

\begin{figure}
\centering
\includegraphics[width= 0.5 \textwidth]{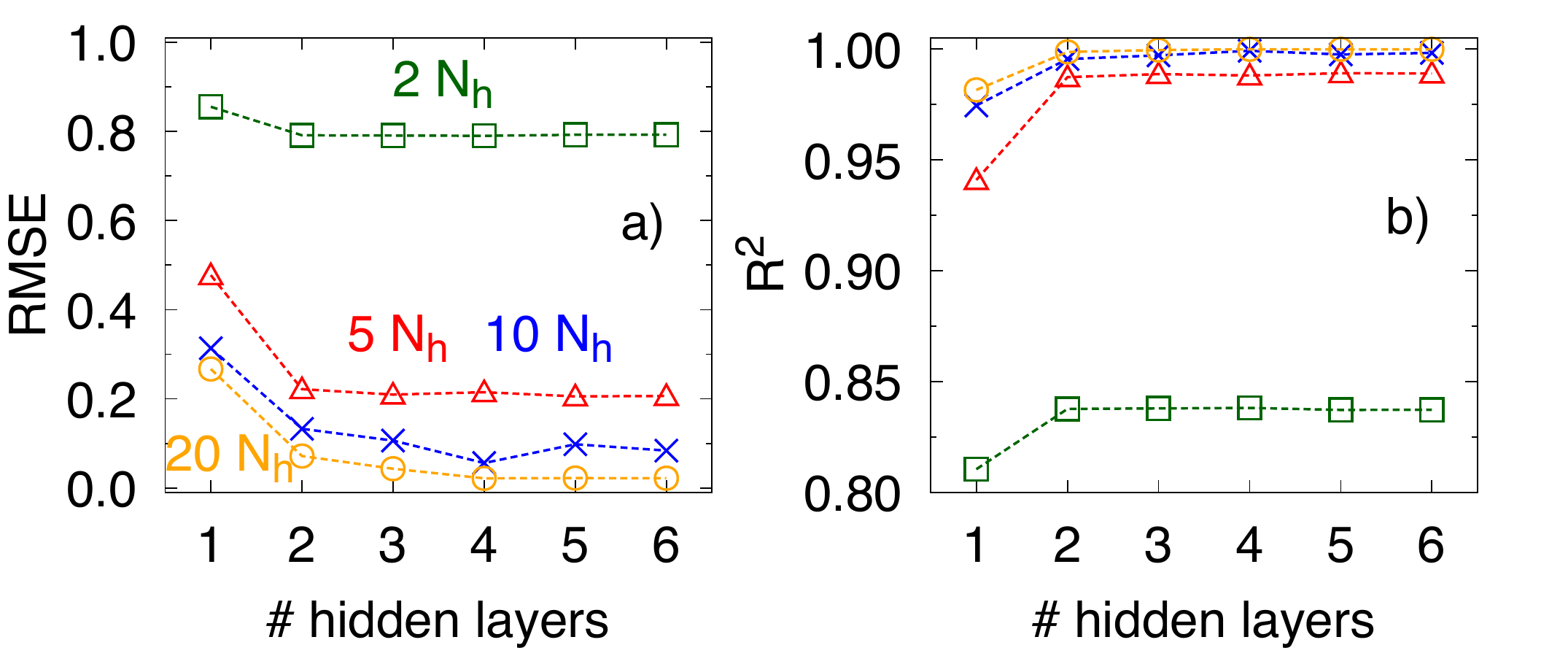}
\caption{Test data $RMSE$ (a) and $R^2$ (b) against number of hidden layers for different number of neurons. Square points represent $2$ hidden neurons ($N_h$), triangle points represent $5N_h$ (5 hidden neurons), cross points represent $10N_h$ and circle points represent $20N_h$. }
   \label{error_2}
\end{figure}

\begin{figure}
\centering
\includegraphics[width= 0.5 \textwidth]{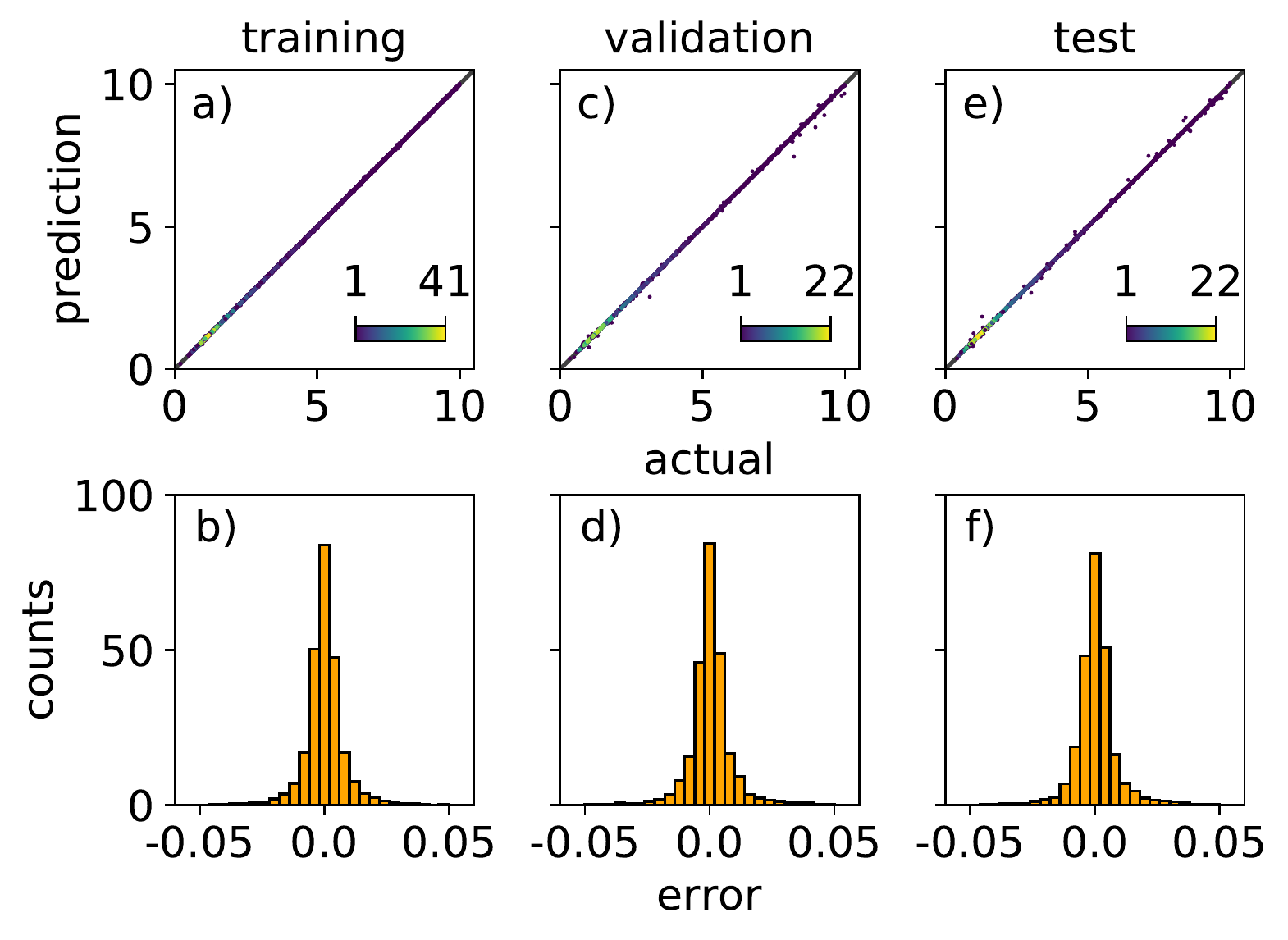}
\caption{Scatter plot between actual and predicted $F$ (top row)  and  error (actual $F$ - predicted $F$) distribution (bottom row) for training (a, b), validation (c, d) and test (e, f) data. }
   \label{pred}
\end{figure}

Selection of network architecture that includes the number of hidden layers, number of hidden neurons, initial $\sigma$, $\sigma_d$, $\sigma_i$, activation functions etc. is very crucial. For very small number of hidden layers and hidden neurons the underlying nonlinear mapping can not be approximated effectively. On the other hand for very big network there is a larger probability of  over-fitting. Therefore the choice of optimal number of hidden layers and hidden neurons is very important. Based on a trial and error method, we find that in our case, a network with  $4$ hidden layers and $20$ neurons  each performs best. Prediction performance also crucially depends on the  value of $\sigma$. We adapt the value of $\sigma$ during the optimization by decreasing or increasing its value depending on the performance function. We tried different combinations of the initial values of $\sigma$, $\sigma_d$ and $\sigma_i$ and it turned out that $\sigma=0.001$, $\sigma_d=0.1$ and $\sigma_i=10$ performed the best.    

We used $70\%$ of the full data ($30000$ pairs) for the training and $15\%$ data each for the validation and the test processes. The convergence of  $RMSE$ and $R^2$ on test data against training size are shown in Fig.(\ref{error}). The error bars are obtained from $10$ random trials. When we increase training size, $RMSE$ ($R^2$) initially decreases (increases) rapidly and then varies very slowly.  After about $10000$ training size, the error saturates on a convergence plateau. For full training data (i.e. $21000$ training data), we calculate the mean absolute error ($MAE$) and the mean absolute percentage error ($MAPE$) using following formulae:
\begin{eqnarray}
\label{mse_r2}
MAE&=&\frac{1}{N}\sum_i |F_i^{true}-F_i^{pred}|, \\
MAPE&=&\frac{1}{N}\sum_i \bigg{|} \frac{F_i^{true}-F_i^{pred}}{F_i^{true}}\bigg{|}\times 100, 
\end{eqnarray}  
along with $RMSE$ and $R^2$ and are listed  in Table.(\ref{error_table}). Note that $MAE$ has the same unit as $F$.

Figure (\ref{error_2}) displays  $RMSE$ and $R^2$ for the test data as a function of number of hidden layers for different number of hidden layer neurons. From Fig.(\ref{error_2}) it is clear that $RMSE$ ($R^2$) decreases (increases) rapidly at first and then varies slowly with hidden layer size. The saturated value of $RMSE$ ($R^2$) decreases (increases) with the number of hidden neurons. $RMSE$ ($R^2$) with $2$ neurons saturates at about $0.7925$ ($0.8373$) whereas for $20$ neurons it saturates at about $0.0235$ ($0.9998$). There is a considerable error improvement when we change the number of neurons from $2$ neurons to $5$ neurons. After this, the improvement is very low and converges after $10$ neurons with a hidden layer size $\geq 3$.

In top row of Fig.(\ref{pred}) we show a scatter plot between the actual (obtained numerically) and the predicted (obtained using the trained ANN) values of  $F$. The solid diagonal line represents the identity line $y=x$. Ideally all points should lie along the diagonal line (which is the case for the best prediction). We observe a good prediction in our case as almost all points are very close to the diagonal line for three cases. Error distributions (actual value - predicted value) are shown in the bottom row of Fig.(\ref{pred}). All error distributions are Gaussian in shape and peaked around zero with very small standard deviation.

\begin{table}
\centering
\caption{$MAE$, $MAPE$, $RMSE$ and $R^2$ for training, validation and test data for $21000$ training size.}
\label{error_table}
\begin{tabular}{ c c c c} 
 \hline
 \hline
& training &  validation & test \\
\hline
 $MAE$ & $0.0054$ & $0.0073$ & $0.0074$ \\ 
 $MAPE(\%)$ & $0.2796$ & $0.3560$ & $0.3610$ \\ 
 $RMSE$ & $0.0087$ & $0.0234$ & $0.0235$ \\ 
 $R^2$ & $0.9999$ & $0.9998$ & $0.9998$ \\ 
 \hline
 \hline
\end{tabular}
\end{table}

\section{Nonequilibrium photon exchange statistics}
\label{stats}
\begin{figure}
\centering
\includegraphics[width= 0.5 \textwidth]{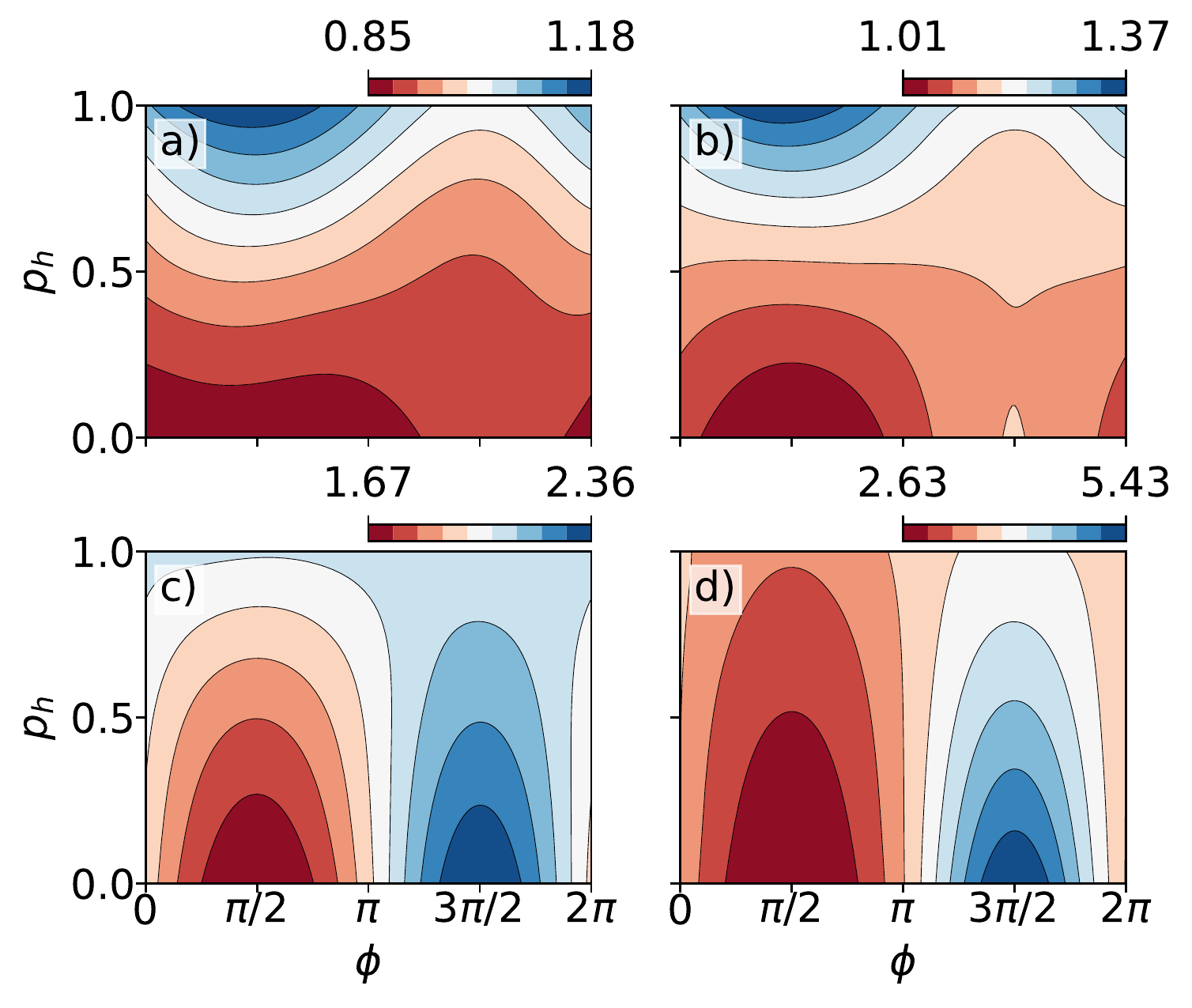}
\caption{Contour plot of $F$ as a function of  $\phi$ and $p_h$ for $T_l=0.1$ (a), $T_l=0.3$ (b), $T_l=0.5$ (c) and $T_l=0.7$ (d). Rest of the parameters are fixed at $T_{c0}=0.6$, $T_{h0}=1.6$ and $p_c=0.5$.}
   \label{fano_contour}
\end{figure}

We are now in a position to analyze the statistics of photon exchange using the trained ANN. We study $F$ as a function of hot bath coherence ($p_h$) and phase difference ($\phi$) between two temperature driving protocols for different cavity temperatures ($T_l$) and are shown in Figs.(\ref{fano_contour}a-d). We see that the two regimes, $F>1$ and $F< 1$, are visible in Fig.(\ref{fano_contour}a), indicating that both coherences and PBp effects can be tuned to generate an interplay between the two statistics, i.e. bunched or super-Poissonian ($F>1$) and antibunched or sub-Poissonian ($F<1$) photon exchange statistics. In  Fig.(\ref{fano_contour}a), we see that as the phase difference (a measure of geometric contributions at $T_l=0.1$) increases, the statistics tend to remain antibunched for a large range of $p_h$ values. Only in a small regime (blue contours at higher values of $p_h$) bunched statistics are obtained. For such small $T_l$ values, $F$ increases with $p_h$ for all values of $\phi$. As we slightly increase $T_l$, the $p_h$ values for which $F<1$ vanishes as seen in Fig.(\ref{fano_contour}b). Thus the statistical  dependence of $F$ on $T_l$ is very sensitive. Even for a slight increase in $T_l$, the antibunched statistics vanish. This is because, an increase in $T_l$ increases the cavity occupation number $n_l$ which is an indication of classical (or thermal) photonic behavior. For even higher values of $T_l$ (Figs.(\ref{fano_contour}c,d)), we obtain some giant Fano factors. Giant Fano factors represent highly bunched photon transfer statistics and have also been  observed in electron transport through single molecules \cite{PhysRevLett.94.206804}. These giant Fano factors are a result of severe suppression of flux (mean) in comparison to the noise.

As the cavity temperature, $T_l$ is further increased, we can see an oscillation in $F$ as a function of $\phi$ as shown in Fig.(\ref{fano_contour}d). In this figure,  $F$ increases with $p_h$ for $0<\phi<\pi$, which then decreases with $p_h$ for $\pi<\phi<2\pi$. This type of oscillation is however absent in  Fig.(\ref{fano_contour}a) where the value of $T_l$ is low. This appearance of oscillation indicates the presence of a pivotal $T_l$ which dictates this oscillation (from lower to giant Fano factors) of $F$ as a function of $\phi$. The reason for this oscillation can be explained as follows. Both the geometric cumulants oscillate as a function of $\phi$ but with different amplitudes and this effect is seen in the total cumulant when the geometric contributions are large enough. As $T_l$ is increased, the amplitude at which the fluctuations oscillate is larger than the amplitude at which the flux oscillates. At lower $T_l$, the magnitudes are almost the same and hence we do not see oscillations. 
We conclude by stating that at lower values of $T_l$ (below the pivotal $T_l$), at fixed values of $p_h$, $F$ does not change much with $\phi$ but for higher value of $T_l$ (above the pivotal $T_l$) it shows an oscillatory behavior  as a function of $\phi$ which results from unequal magnitudes of oscillation in the flux and fluctuations.

\section{Thermodynamic Uncertainty Relationship}
\label{tur}
The thermodynamic uncertainty relationship is one of the recent developments in nonequilibrium thermodynamics \cite{pietzonka2017finite} and represents a fundamental relation between the nonequilibrium fluctuation (variance) and the entropic cost. It has been theoretically proven in several systems ranging from biomolecules \cite{barato2015thermodynamic}, quantum junctions\cite{agarwalla2018,PhysRevB.98.085425} to periodically driven systems\cite{koyuk2018generalization}. It has also been claimed to be universally valid for Markovian dynamics in several systems \cite{barato2015thermodynamic, pietzonka2017finite,proesmans2017discrete,horowitz2017proof}.
As per the thermodynamic uncertainty relationship, the entropy production rate, $\dot \Sigma$ obeys the equation 
\begin{align}
 \frac{C^{(2)}}{(C^{(1)})^2}\dot \Sigma \ge 2k_B.
\end{align}
Here, $k_B$ stands for the Boltzmann constant. From 
the principles of quantum transport, $\dot\Sigma = C^{(1)}{\cal A}$ \cite{agarwalla2018}, where ${\cal A}$ is a thermodynamic affinity and can be obtained from a steadystate fluctuation theorem, 
\begin{align}
\label{ssft}
 \frac{P(q,t)}{P(-q,t)}=e^{q{\cal A}}.
\end{align}
It is now straightforward to show that ($k_B=1$), 
\begin{align}
 F{\cal A}\ge 2.
\end{align}
It has been shown by us previously that in the dQHE the affinity is given by\cite{PhysRevE.96.052129} 
 \begin{eqnarray}
 {\cal A} = \ln \frac{\tilde n_l
 \int_0^{t_p}(1+ n_c(t')) n_h(t')dt'}{n_l\int_0^{t_p}n_c(t')(1+ n_h(t'))dt'},
 \end{eqnarray}
which in absence of geometric contributions, satisfies the steadystate fluctuation theorem  given by Eq.(\ref{ssft}).

We report that as long as the geometric contributions are zero, the thermodynamic uncertainty holds irrespective of any quantum coherence  values. This is in contrast to what has been reported in some works where small regimes in the coupling and coherence parameter space exists such that the inequality is broken\cite{PhysRevB.98.085425, agarwalla2018}. It has been attributed to the fact that one can decrease fluctuations by tuning coherences and couplings to obtain $F<1$ such that the uncertainty relation is violated. However, we note that although such regimes can be achieved in our model, we find that a trade-off between flux and fluctuations  ensures the validity of the the uncertainty relationship. One can argue that the affinity doesn't depend on the coherences  hence the regimes with lower value ${\cal A}$ can be identified to break the inequality. However, based on the values obtained from our ANN,  we find that even for such a parameter scenario, $ F{\cal A}\ge 2$ holds. We find that the cases with low values of ${\cal A}$ are compensated by a drastic increase in $F$. This happens because low thermodynamic affinities result in a lower value of the flux and higher values of fluctuations giving rise to giant Fano factors. Likewise, higher affinities result in higher values of flux giving smaller values of $F$. This trade-off physically {\em represents} the thermodynamic uncertainty.   Note that the couplings are assumed to be equal in our case and hence we have definitely missed out a large chunk of the parameter space which do not allow us to concretely confirm whether coherences can break the above uncertainty relation in the unequal coupling regime. But in the equal coupling, the thermodynamic uncertainty holds and cannot be broken in presence of coherences. This is also true for a range of couplings that we tested, $0<r,g\le 10$.

\begin{figure}[t]
\centering
\includegraphics[width= 0.5 \textwidth]{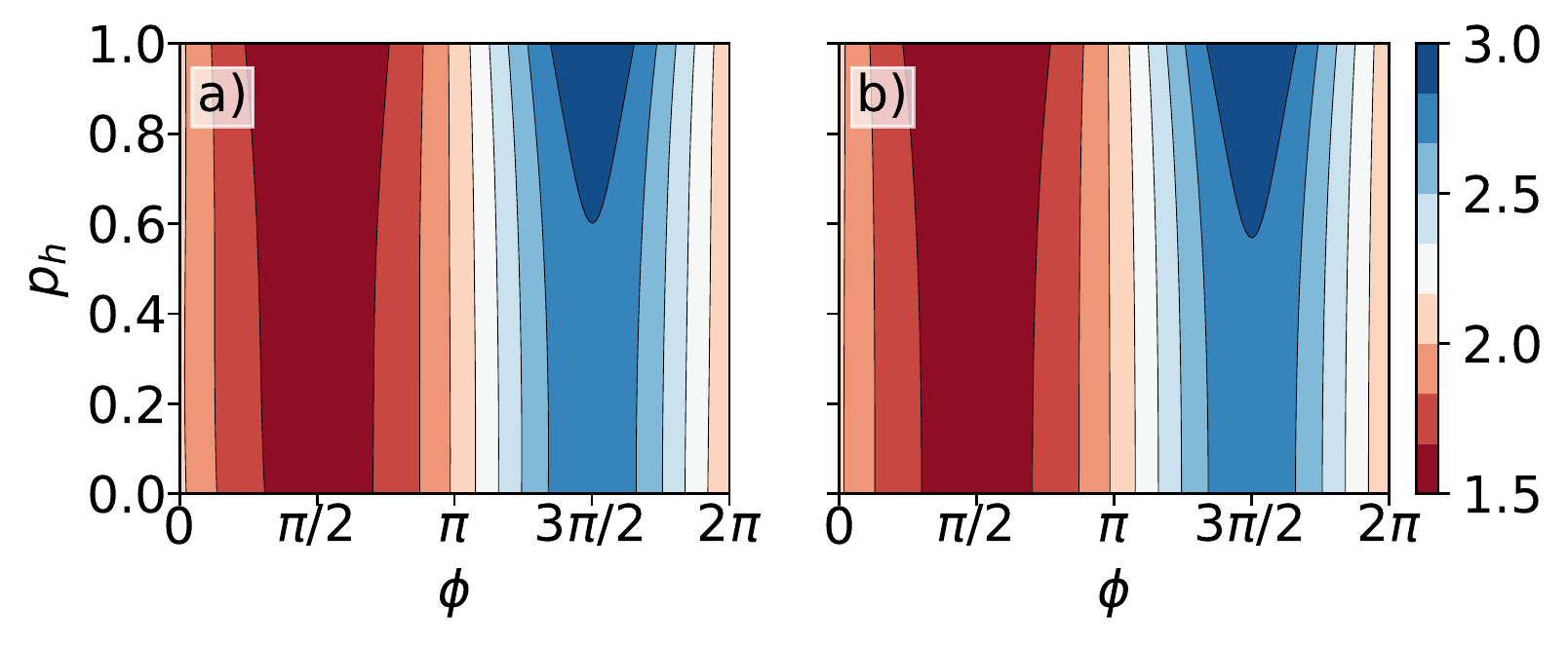}
\caption{Contour plot of $FA$ against $\phi$ and $p_h$ using the trained network (a) and solving the quantum master equation (b). Other parameters are fixed at $T_{c0}=0.6$, $T_{h0}=1.6$, $T_l=0.7$ and $p_c=0.0$.}
   \label{thermo_uncert}
\end{figure}

The above arguments, however, cannot be regarded as true in presence of finite geometric effects and the thermodynamic uncertainty relationship doesn't hold. In Fig.(\ref{thermo_uncert})  the quantity $F{\cal A}$ contour is shown as a function of $p_h$ and $\phi$ using values obtained from the trained network (Fig.(\ref{thermo_uncert},a)) as well as numerical simulation (Fig.(\ref{thermo_uncert},b)). Both the figures are strikingly similar and further establishes the good performance of the ANN. As can be seen, $F{\cal A}>2$ is fully maintained along the $p_h$ axis at $\phi = 0, \pi$ and $2\pi$ showing that the inequality holds as a function of coherences. However as a function of $\phi$, $F{\cal A}<2$ appears within $0<\phi<\pi$ where the geometric contributions severely increases the total flux giving low values of the Fano factors violating the mathematical uncertainty. We point out that this is solely because of the phase difference and not due to the coherences since at $0,\pi,2\pi$, all $p_h$ values preserve the uncertainty.

The physical explanation of the invalidity of the thermodynamic uncertainty relationship has its origins rooted in the steadystate fluctuation theorem. In presence of finite geometric effects, one cannot write down or derive a standard form the fluctuation theorem shown in Eq.(\ref{ssft}). The mathematical breakdown for the thermodynamic uncertainty relationship is a direct consequence of the fact that the fluctuation theorem is violated in presence of PBp effects. With finite PBp effects, ${\cal A}$ no longer represents  a  correct thermodynamic affinity since Eq.(\ref{ssft}) is violated. The actual effective affinity that drives the system out of equilibrium needs to be reformulated by including geometric corrections to it. Unfortunately, there is no straight forward way to  determine ${\cal A}$ within the present formalism and hence we leave it as a future direction of research. Infact, a proper theory that explains the thermodynamics in presence of geometric effects is lacking in the literature. We conclude by stating that the invalidity of the thermodynamic uncertainty relationship is due to the violation of the steadystate fluctuation theorem in presence of finite geometric effects. It however holds true  when geometric contributions are zero.

\section{Conclusion}
\label{conc}
In this work, we  explored the dependence of system parameters on the non equilibrium photon exchange statistics between  system and cavity of an adiabatically temperature-driven quantum heat engine using an artificial neural network via regression analysis. We found a pivotal cavity temperature beyond which there exists a low to giant Fano factor oscillation as a function of phase difference. This oscillation results from unequal magnitudes of the oscillation in the steadystate cumulants as a function of the phase difference between the driving protocols. We also show that both geometric phase and coherences can be used to alter the statistics of photon exchange from bunched to antibunched. The antibunched statistics disappear at larger cavity temperatures and cannot be observed even in presence of large geometric contributions. We further show that the thermodynamic uncertainty relationship doesn't hold true in presence of geometric effects. The  breakdown of the uncertainty relationship is attributed to the violation of the steadystate fluctuation theorem in presence of geometric effects. When the phase difference is an integral multiple of $\pi$, the geometric contributions vanish and the uncertainty relationship holds since the fluctuation theorem is recovered. We also report that the uncertainty relationship is robust against quantum coherences, atleast in the equal coupling limit.

\begin{acknowledgments}
 We acknowledge the support from the Max-Planck-Institute for the Physics of Complex Systems, Dresden, Germany where all of the work was carried out. HPG also acknowledges the support from the Department of Chemical Sciences, Tezpur University, where he is currently affiliated.
 HPG would also like to express his immense gratitude and respect for  his recently expired mother, Lalita Devi Goswami for her never ending emotional support at all times.
 \end{acknowledgments}

\bibliography{referencesArXiv.bib}

  \end{document}